\documentclass[iop,numberedappendix]{emulateapj}

\slugcomment{Draft version \today}

\shorttitle{Stellar Multiplicity in APOGEE}
\shortauthors{C. Badenes et al.}

\usepackage{graphicx}	
\usepackage{amsmath}	
\usepackage{amssymb}	
\usepackage[dvipsnames]{xcolor}
\usepackage{hyperref}
\hypersetup{
	colorlinks   = true, 
	urlcolor     = blue, 
	linkcolor    = blue, 
	citecolor   = blue 
}

\usepackage{float}

\newcommand{\msun}{M$_\odot$}

\newcommand{\kms}{\,{\rm km \, s$^{-1}$}}

\newcommand{\drvm}{$\Delta RV_{max}$}
\newcommand{\drvpp}{$\Delta RV_{pp}$}
\newcommand{\teff}{$T_{eff}$}
\newcommand{\logg}{$\log g$}
\newcommand{\logp}{$\log P$}

\begin{document}


\title{Stellar Multiplicity Meets Stellar Evolution and Metallicity: The APOGEE View}


\author{Carles Badenes\altaffilmark{1,2}, Christine Mazzola\altaffilmark{1}, Todd A. Thompson\altaffilmark{3}, Kevin
  Covey\altaffilmark{4}, Peter E. Freeman\altaffilmark{5}, Matthew G. Walker\altaffilmark{6}, Maxwell Moe\altaffilmark{7,8},
  Nicholas Troup\altaffilmark{9}, David Nidever\altaffilmark{10}, Carlos Allende Prieto\altaffilmark{11}, Brett Andrews\altaffilmark{1}, Rodolfo
  H. Barb\'a\altaffilmark{12}, Timothy C. Beers\altaffilmark{13}, Jo Bovy\altaffilmark{14,15}, Joleen K. Carlberg\altaffilmark{16}
  Nathan De Lee\altaffilmark{17,18}, Jennifer Johnson\altaffilmark{3}, Hannah Lewis\altaffilmark{9}, Steven
  R. Majewski\altaffilmark{9}, Marc Pinsonneault\altaffilmark{3}, Jennifer Sobeck\altaffilmark{19}, Keivan
  G. Stassun\altaffilmark{18}, Guy S. Stringfellow\altaffilmark{20}, Gail Zasowski\altaffilmark{21}}


\altaffiltext{1}{Department of Physics and Astronomy and Pittsburgh Particle Physics, Astrophysics and Cosmology Center (PITT
  PACC), University of Pittsburgh, 3941 O'Hara Street, Pittsburgh, PA 15260, USA, \email{badenes@pitt.edu}}
\altaffiltext{2}{Institut de Ci\`encies del Cosmos (ICCUB), Universitat de Barcelona (IEEC-UB), Mart\'i Franqu\'es 1, E08028 Barcelona, Spain}
\altaffiltext{3}{Department of Astronomy and the Center for Cosmology and Astro-Particle Physics, The Ohio State University, Columbus, OH 43210, USA}
\altaffiltext{4}{Department of Physics and Astronomy, Western Washington University, Bellingham, WA, 98225, USA}
\altaffiltext{5}{Department of Statistics, Carnegie Mellon University, 5000 Forbes Avenue, Pittsburgh, PA 15213, USA}
\altaffiltext{6}{McWilliams Center for Cosmology, Department of Physics, Carnegie Mellon University, 5000 Forbes Avenue, Pittsburgh, PA 15213, USA}
\altaffiltext{7}{Steward Observatory, University of Arizona, 933 N. Cherry Ave., Tucson, AZ 85721, USA}
\altaffiltext{8}{\textit{Einstein} Fellow}
\altaffiltext{9}{Department of Astronomy, University of Virginia, Charlottesville, VA 22904-4325, USA}
\altaffiltext{10}{National Optical Astronomy Observatory, 950 North Cherry Avenue, Tucson, AZ 85719, USA}
\altaffiltext{11}{Instituto de Astrof\'isica de Canarias, V\'ia L\'actea, 38205 La Laguna, Tenerife, Spain}
\altaffiltext{12}{Departamento de F\'isica y Astronom\'ia, Av. Juan Cisternas 1200 Norte, Universidad de La Serena, La Serena,
  Chile}
\altaffiltext{13}{Department of Physics and JINA Center for the Evolution of the Elements, University of Notre Dame, Notre Dame, IN  46556, USA}
\altaffiltext{14}{Department of Astronomy and Astrophysics and Dunlap Institute for Astronomy and Astrophysics, University of
  Toronto, 50 St. George Street, Toronto, Ontario M5S 3H4, Canada}
\altaffiltext{15}{Alfred P. Sloan Fellow}
\altaffiltext{16}{Space Telescope Science Institute, 3700 San Martin Dr., Baltimore MD 21218}
\altaffiltext{17}{Department of Physics, Geology, and Engineering Tech, Northern Kentucky University, Highland Heights, KY 41099,
  USA}
\altaffiltext{18}{Department of Physics and Astronomy, Vanderbilt University, Nashville, TN, USA}
\altaffiltext{19}{Department of Astronomy, Box 351580, University of Washington, Seattle, WA 98195, USA}
\altaffiltext{20}{Center for Astrophysics and Space Astronomy, Departmentnof Astrophysical and Planetary Sciences, University of
  Colorado, Boulder, CO, 80309, USA}
\altaffiltext{21}{Department of Physics and Astronomy, University of Utah, Salt Lake City, UT, 84112, USA}


\begin{abstract}
  We use the multi-epoch radial velocities acquired by the APOGEE survey to perform a large scale statistical study of stellar
  multiplicity for field stars in the Milky Way, spanning the evolutionary phases between the main sequence and the red clump. We
  show that the distribution of maximum radial velocity shifts (\drvm) for APOGEE targets is a strong function of \logg, with main
  sequence stars showing \drvm\ as high as $\sim$300 \kms, and steadily dropping down to $\sim$30 \kms\ for \logg$\sim$0, as stars
  climb up the Red Giant Branch (RGB). Red clump stars show a distribution of \drvm\ values comparable to that of stars at the tip
  of the RGB, implying they have similar multiplicity characteristics. The observed attrition of high \drvm\ systems in the RGB is
  consistent with a lognormal period distribution in the main sequence and a multiplicity fraction of 0.35, which is truncated at
  an increasing period as stars become physically larger and undergo mass transfer after Roche Lobe Overflow during H shell
  burning. The \drvm\ distributions also show that the multiplicity characteristics of field stars are metallicity dependent, with
  metal-poor ([Fe/H]$\lesssim-0.5$) stars having a multiplicity fraction a factor 2-3 higher than metal-rich ([Fe/H]$\gtrsim0.0$)
  stars. This has profound implications for the formation rates of interacting binaries observed by astronomical transient surveys
  and gravitational wave detectors, as well as the habitability of circumbinary planets.
\end{abstract}





\section{Introduction} \label{Introduction}

\begin{figure}
\centering
\includegraphics[width=0.48\textwidth]{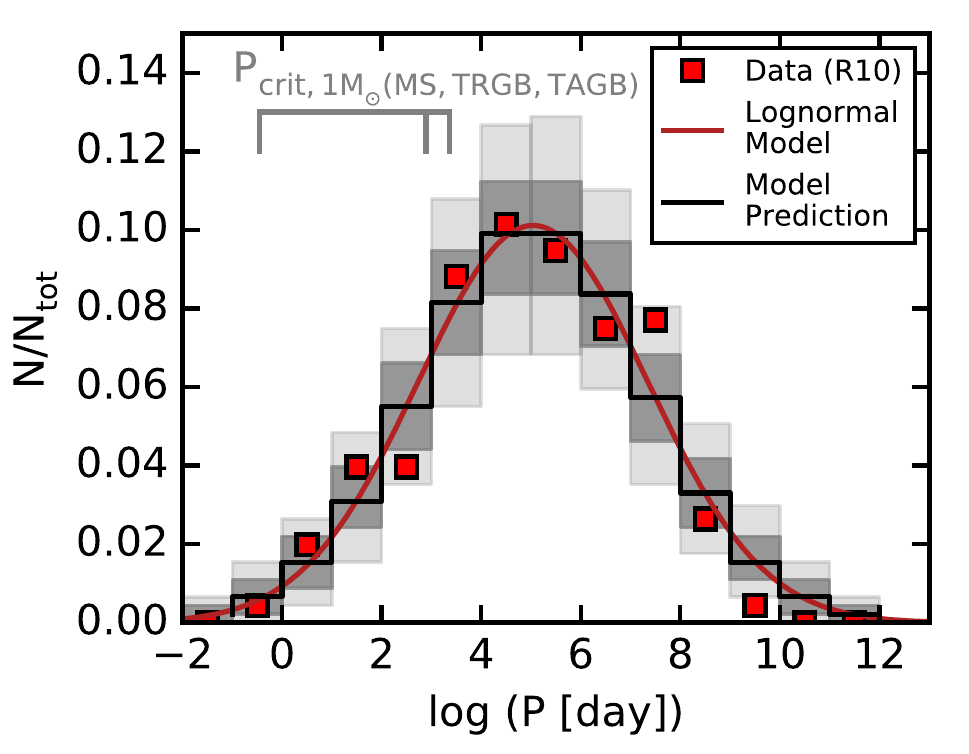}
\caption{Period distribution for the sun-like MS stars in \cite{Raghavan2010} (red squares), together with a model lognormal
  distribution ($\overline{\log{P}} = 5.0$ and $\sigma_{\log{P}} = 2.3$, dark red solid line), and the predicted nominal
  (black), and 1 and 2$\sigma$ ranges (dark and light gray) from Poisson realizations of the model. The ruler in the top left
  corner indicates the critical period for Roche Lobe overflow (RLOF) in Sun-like models during the MS, at the tip of the Red
  Giant Branch (RGB), and at the tip of the Asymptotic Giant Branch (AGB).}
\label{RaghPDist}
\end{figure}

Stellar multiplicity plays a fundamental role in astrophysics. Interacting binary stars are the progenitors of all Type Ia SNe
\citep{Maoz2014}, many core collapse SNe \citep{Sana2012}, and a host of other astronomical sources, from high- and low-mass X-ray
binaries to novae, cataclysmic variables, AM CVn stars, and most stellar sources of gravitational waves. Yet, our knowledge of the
fundamental statistics of stellar multiplicity (multiplicity fraction, period distribution, mass ratio distribution, and
eccentricity distribution) is still rudimentary, especially after the main sequence (MS) and beyond the Solar neighborhood
\citep[see][for recent reviews]{Duchene2013,Moe2017}. Solar-type MS stars closer than 25 pc have a roughly lognormal period
distribution, with $\overline{\log{P}}\sim5.0$ and $\sigma_{\log{P}} \sim 2.3$ for $P$ in days (\citealt{Raghavan2010}, see Figure
\ref{RaghPDist}), and a multiplicity fraction of $0.50\pm0.04$ for \logp$\leq8$ after completeness corrections \citep{Moe2017}
\footnote{Although it remains the best studied and most comprehensive census of multiple systems in the Solar neighborhood, the
  \cite{Raghavan2010} sample is known to be incomplete for faint companions and very low mass ratios, especially at wide
  separations, see \cite{Chini2014} and \cite{Moe2017} for discussions.}. The multiplicity fraction on the MS is a strong function
of primary mass, with more massive primaries having a higher likelihood to be in a multiple system \citep{Lada2006,Duchene2013},
especially at short periods \citep{Moe2017}, but the lognormal shape of the period distribution is robust below \logp$\sim4$ for
Sun-like stars \citep{Latham2002,Melo2003,Carney2005,Geller2012,Leiner2015,Moe2017}. The mass ratio has a roughly flat distribution, but
\cite{Moe2017} showed that this is not independent of the period distribution. The eccentricities of MS binaries follow a
Maxwellian ``thermal'' distribution at intermediate periods, but tidal interactions circularize the orbits below \logp$\sim1.1$
\citep{Zahn1989,Raghavan2010,Moe2017}.

Most observational studies of stellar multiplicity have focused on small samples of a few hundred objects in specific environments
like the Solar neighborhood or individual stellar clusters
\citep[e.g.][]{Duquennoy1991,Carney2003,Geller2008,Raghavan2010,Matijevic2011,Sana2012,Merle2017}. In this context, intrinsic
variations of the multiplicity statistics with parameters like effective gravity, age, metallicity, or disk/halo membership cannot
be explored without addressing large observational biases. These intrinsic variations, if present, might affect the interplay
between stellar multiplicity and stellar evolution, and impact the formation rates of interacting binaries
\citep{Gosnell2015,Andrews2016}. There is a pressing need for multiplicity surveys with enough scope and precision to deal with
these issues and effectively probe stellar multiplicity after the MS and across the Milky Way.

The Apache Point Observatory Galactic Evolution Experiment \citep[APOGEE][]{Majewski2017} within the Sloan Digital Sky Survey
\citep[SDSS,][]{Gunn2006,Blanton2017}, with its high-resolution, high-efficiency, multiplexed infrared spectrograph, has produced
the first truly panoramic view of the stellar content of our Galaxy. APOGEE has targeted more than 150,000 stars at distances of
up to $\sim$30 kpc, probing heavily obscured parts of the Galactic Disk. The APOGEE Stellar Parameter and Chemical Abundances
Pipeline \citep[ASPCAP,][]{GarciaPerez2016} provides reliable measurements of effective temperatures (\teff), surface gravities
(\logg) and chemical abundances for each of these targets, as well as highly precise radial velocities
\citep[RVs,][]{Nidever2015}. APOGEE also has a temporal dimension, with multiple spectra taken for each target, which enables
studies of stellar multiplicity. In \cite{Troup2016}, this temporal dimension was explored for a few thousand stars with seven or
more RV measurements, which allowed to derive orbital solutions and investigate the propertiers of stellar and substellar
companions in detail. Here we focus on the majority of APOGEE targets, which have only sparsely sampled RV curves.

For these stars, detailed orbital parameters cannot be measured \citep[see][]{Price-Whelan2017}, but a statistical analysis of
their RVs can still yield valuable insights on stellar multiplicity, its dependence on fundamental stellar properties, and its
interplay with stellar evolution. Our work continues the statistical study of stellar multiplicity with sparsely sampled RV curves
in SDSS that began with the study of white dwarfs \citep{Badenes2012,Maoz2012} and MS stars \citep{Hettinger2015} drawn
from the low-resolution optical spectra in the Sloan Extension for Galactic Understanding and Exploration survey
\citep[SEGUE,][]{Yanny2009}. Here, we explore the possibilities for these kind of studies that are opened up by the higher
resolution of the infrared APOGEE spectrographs. Our work is organized as follows. In Section \ref{sec:sample-selection} we
describe our sample selection. In Section \ref{sec:drvm:-figure-merit} we examine the statistical properties of the RV variability
in APOGEE stars, its theoretical interpretation, and the dependence with \logg\ and metallicity. In Sections \ref{sec:disc} and
\ref{sec:concl} we discuss our results and present our conclusions.

\section{Sample selection and RV Measurements}
\label{sec:sample-selection}

The DR13 APOGEE \texttt{allStar} file contains measurements for 163278 targets \citep{Albareti2016}. From this sample, we removed
all stars flagged as bad \citep[STAR\_BAD set in the ASPCAP flag bitmask,][]{Holtzman2015} and stars targeted as telluric
calibrators \citep[bit 9 set in the \texttt{apogee\_target2} mask,][]{Zasowski2013}. To restrict our measurements to the field, we
also removed stars targeted as cluster members (bit 9 in the \texttt{apogee\_target1} mask and bit 10 in the
\texttt{apogee\_target2} mask). To further ensure a well-characterized sample, we removed stars that did not have acceptable
uncalibrated values of \teff\ and \logg\ from ASPCAP. We did not use the values calibrated with asteroseismology because these are
only available for giant stars, and we are interested in all the stellar evolution phases sampled by APOGEE. This calibration
shifts the surface gravities for giants by $\sim -0.2$ dex, which is not critical for our goals. This left 122141 entries, for
which we examined the individual RVs in the \texttt{allVisit} file. We used only visit spectra that were deemed of high enough
quality to contribute to the combined APOGEE spectrum \citep[the VISIT\_PK indices,][]{Holtzman2015,Nidever2015}, and we further required
a S/N of at least 40 in each of the visit spectra. A total of 91246 unique stars in APOGEE have two or more RVs that pass these
quality cuts.

\begin{figure*}
\centering
\includegraphics[width=0.7\textwidth]{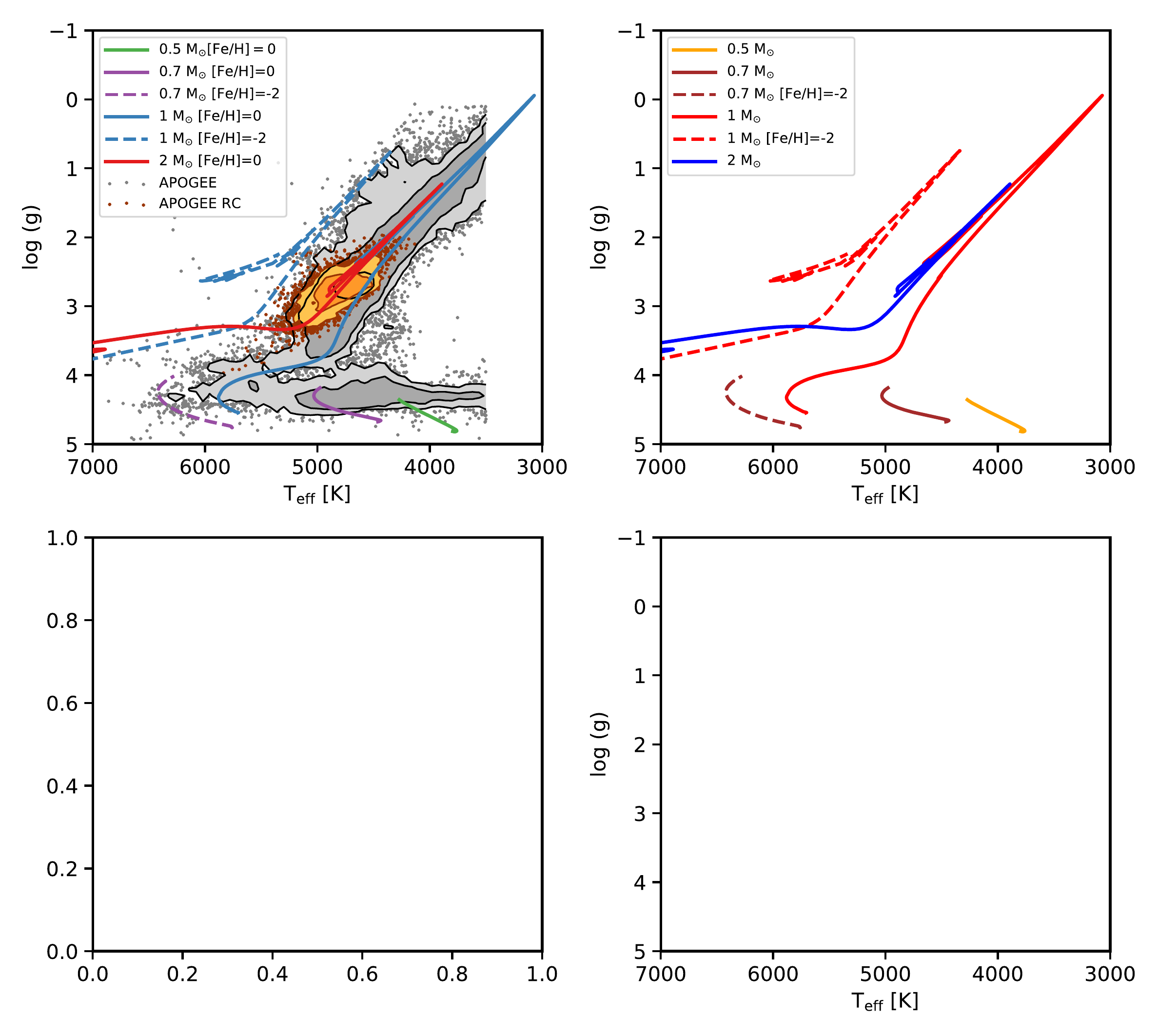}
\caption{Observational HR diagram for the APOGEE targets in our main sample (gray dots and density contours), including the RC stars from
  \cite{Bovy2014} (orange dots and density contours). Selected MIST models from \cite{Choi2016} are overlaid for context.}
\label{HR}
\end{figure*} 

The stars in our main sample are placed in an observational HR diagram in Figure~\ref{HR}. For context, we have overlaid stellar
models from the MESA Isochrone and Stellar Tracks collaboration \citep[MIST,][]{Choi2016,Paxton2011,Paxton2013,Paxton2015},
spanning the phases between core H ignition and core He exhaustion (beginning of the MS to end of the horizontal branch) for
Zero-Age Main Sequence masses between 0.5 and 2 \msun. The sample is dominated by RGB stars at several stages of their evolution,
from the subgiant branch to almost the tip of the RGB ($3.75\gtrsim$\logg$\gtrsim$0). There are also substantial contributions
from MS stars and subgiants (\logg$\gtrsim$ 3.75 and \teff\ between 3500 and 6500 K), as well as Red Clump (RC) stars,
which can be identified by their location in color-metallicity-\logg -\teff\ space \citep{Bovy2014}. The 15667 RC stars from the
DR13 version of the \cite{Bovy2014} catalog that pass our quality cuts are shown separately in Figure~\ref{HR}. Our main sample
probably contains some AGB stars, which overlap the RGB in \logg, \teff\ space, but given the short lifetimes of all phases of
stellar evolution after core He exhaustion, this contamination should be small and we ignored it in our analysis.

\begin{figure}
\centering
\includegraphics[width=0.48\textwidth]{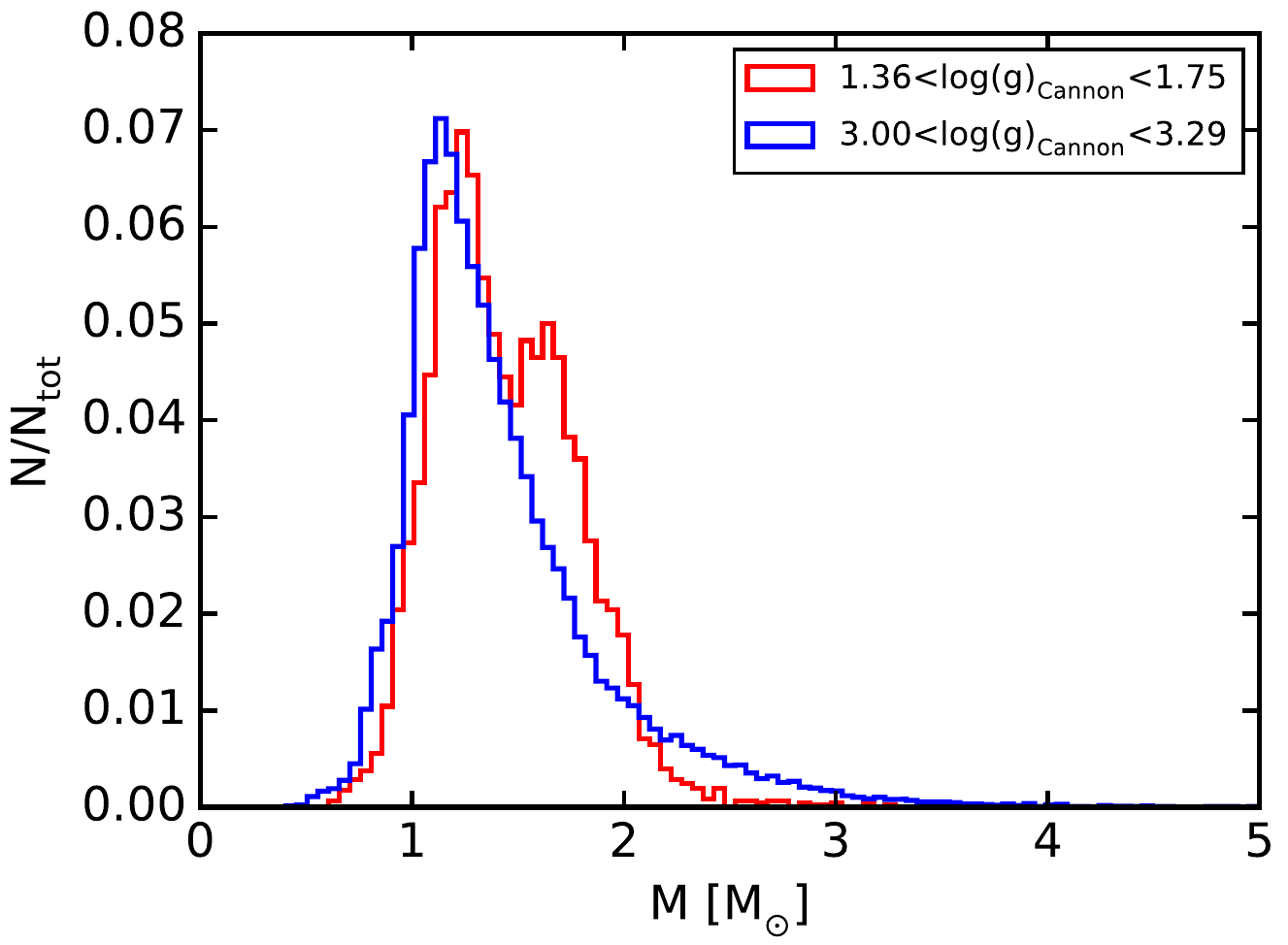}
\caption{Mass distribution of APOGEE targets classified as RGB stars in two representative \logg\ ranges from the re-analysis of
  the spectra by \cite{Ness2016}.}
\label{NessMassHist}
\end{figure}

\begin{figure}
\centering
\includegraphics[width=0.48\textwidth]{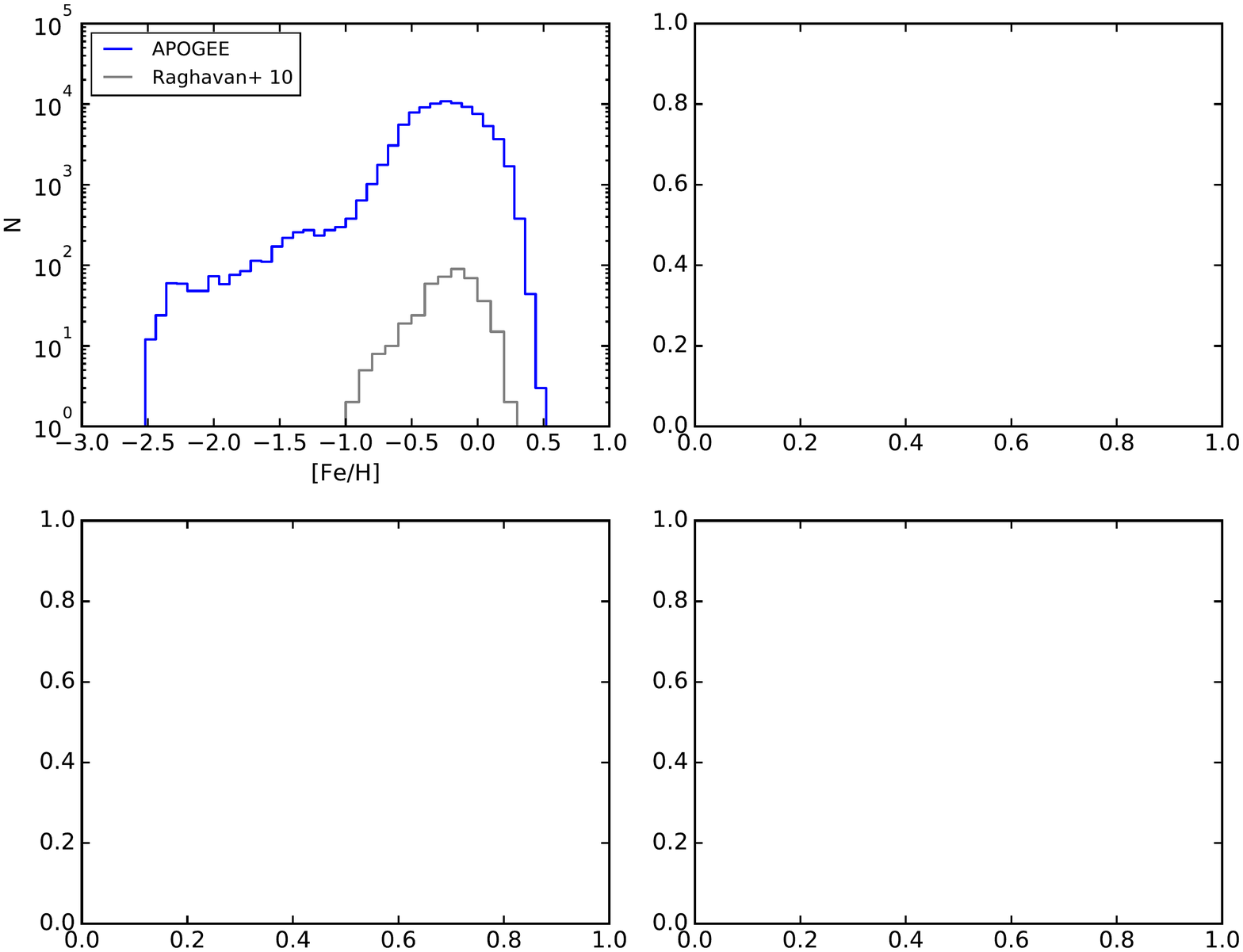}
\caption{Distribution of [Fe/H] values in the APOGEE targets that pass our quality cuts (blue histogram), compared to the sample
  from \cite{Raghavan2010} (gray histogram).}
\label{FeHDist}
\end{figure}

The properties of the stars in our main sample are typical of APOGEE targets with high quality spectra. The metallicities range
from [Fe/H] = 0.5 to -2.5 (the range of models available in ASPCAP), with the bulk of the targets between 0.0 and -0.5 (see
Figure~\ref{FeHDist}). \cite{Ness2016} derived ages and masses for some of the RGB and RC stars in APOGEE by applying the label
transfer tool \textit{The Cannon} \citep{Ness2015} to a training set of stars with asteroseismic masses from APOKASC
\citep{Pinsonneault2014}. This study probably yields the most accurate measurements for the masses and ages of APOGEE targets, but
it does not extend to \logg\ values above 3.29 or below 1.36. In Figure \ref{NessMassHist} we show the mass histograms for the RGB
stars with the highest and lowest \logg\ values in the study by \cite{Ness2016} - the RC mass distribution is omitted for clarity,
but it is similar to that at the highest \logg\ range. From these distributions, we conclude that most RGB and RC stars in APOGEE
have masses close to 1 \msun, with few below that value or above 2 \msun. The mass distribution in the MS is not as well
characterized, but it must be skewed towards lower masses (see Figure \ref{HR}).

The APOGEE data reduction pipeline \citep{Nidever2015} and the ASPCAP pipeline \citep{GarciaPerez2016} operate on the assumption
that each spectrum is dominated by flux emitted by a single stellar photosphere, but this is not necessarily true. Some stars will
have companions of comparable brightness, such that the combined spectrum will reveal spectral lines from both components, often
offset in wavelength according to their relative line of sight velocities.  These so called double-lined spectroscopic binaries,
or SB2s, can confound the APOGEE and ASPCAP pipelines, potentially introducing substantial errors into the stellar parameters
(e.g., \teff, \logg, [Fe/H], etc., see \citealt{ElBadry2018}) or RVs that are inferred from the spectrum. To account for these effects, we drew from a
preliminary catalog of ~1200 SB2s in the DR13 dataset, constructed using techniques developed to identify and characterize SB2s in
the APOGEE/IN-SYNC survey of local star forming regions \citep{Fernandez2017}.  Only 482 of these SB2 stars pass the quality cuts
used to construct our sample, most of them with \logg\ values in the MS and subgiant region. For these targets, we replace the RVs
calculated by the APOGEE pipeline with the RV extracted from the highest peak in the cross-correlation function at each epoch,
which provides a more reliable measure of the RV of the photometric primary. A similar study using a data-driven spectral fitting
method for the MS was recently published by \cite{ElBadry2017}.

\begin{figure*}
\centering
\includegraphics[width=0.7\textwidth]{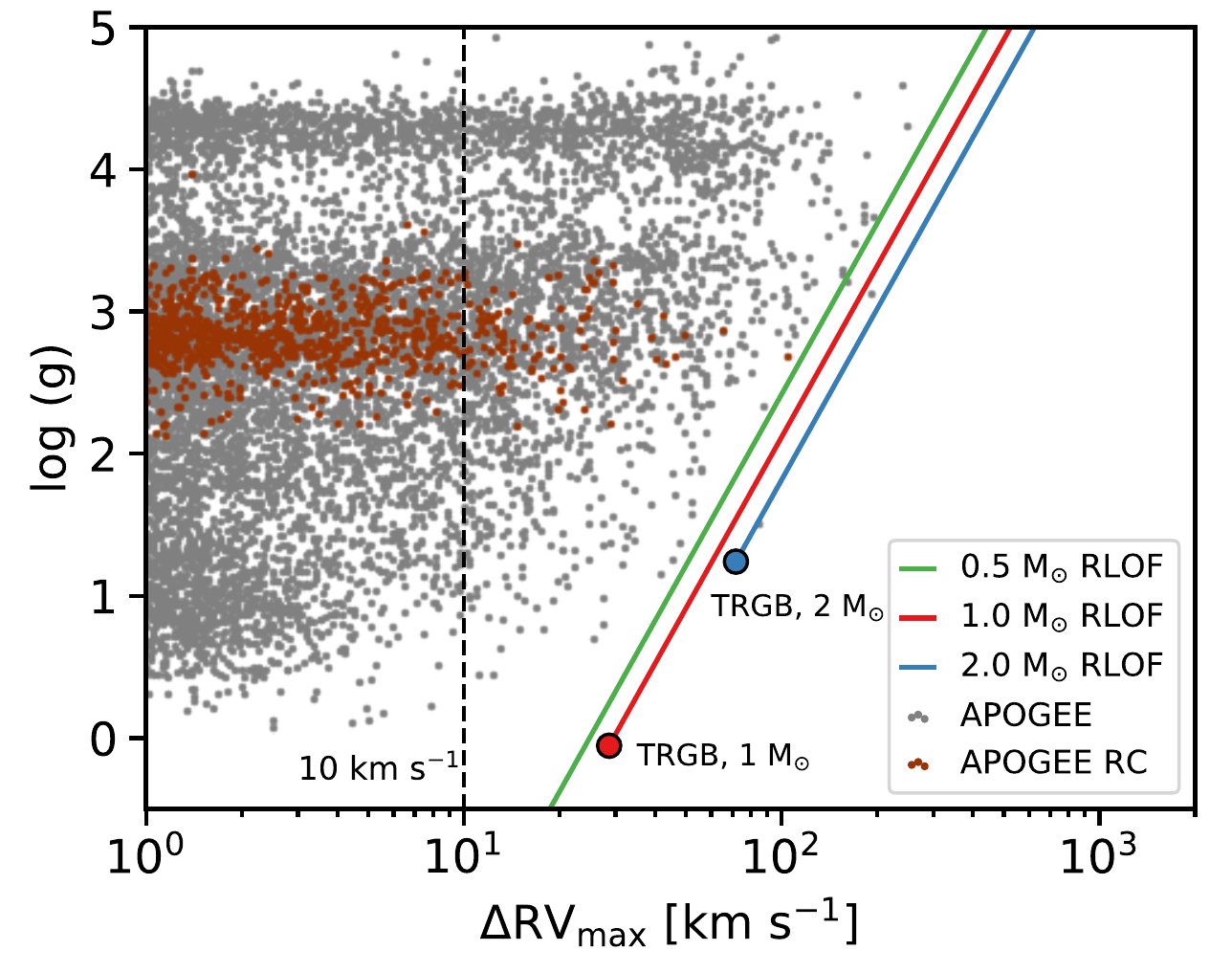}
\caption{Distribution of \drvm\ values for APOGEE stars in our main sample (gray dots) and RC sample (dark red dots) as a function
  of \logg. The solid lines indicate the maximum value of \drvm\ (for $q=1$ and $ i=90^{\circ}$) at the critical period for RLOF
  in stars of 0.5 (green), 1 (red), and 2 (blue) \msun\ as a function of \logg. The position of the tip of the RGB (lowest value
  of \logg) in MIST models of solar metallicity is indicated by the terminal symbols for 1 and 2 \msun\ stars.}
\label{DRVMvlogg}
\end{figure*}

\section{\drvm: a Figure of Merit for Sparsely Sampled RV Curves}
\label{sec:drvm:-figure-merit}

\subsection{\drvm\ and \logg}

Among the targets in our sample, most (62\%) have three RVs that pass quality cuts, 22\% have only 2 RVs, and 16\% have 4 or
more. A simple figure of merit to characterize these sparsely sampled RV curves is the difference between the highest and lowest
measured RVs for each star, \drvm$=\max(\mathrm{RV}_{n}) - \min(\mathrm{RV}_{n})$ \citep[see][for 
discussions]{Badenes2012,Maoz2012}. In Figure~\ref{DRVMvlogg} we show the values of \drvm\ as a function of \logg\ for the targets
in our main sample. The maximum value of \drvm\ measured by APOGEE is a strong function of \logg, with MS stars showing values as
high as $\sim$300 \kms, and stars near the tip of the RGB only going to $\sim$30 \kms. 

\begin{figure}
\centering
\includegraphics[width=0.48\textwidth]{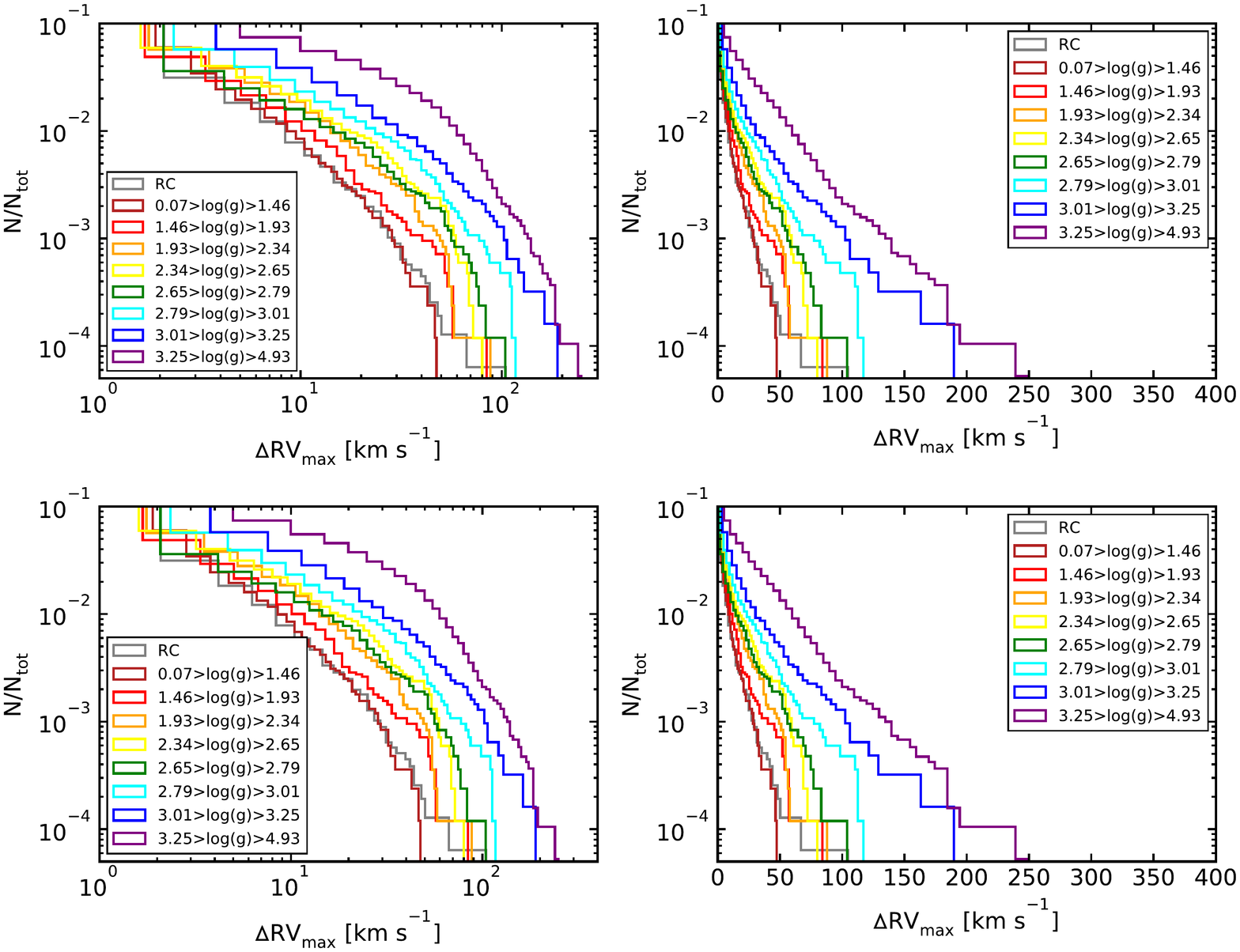}
\caption{Cumulative histograms of \drvm\ for eight \logg\ bins in the main sample, covering the range between the MS and the tip
  of the RGB  (colored plots), plus the RC catalog (gray plot).}
\label{DRVMvloggHist}
\end{figure}

This trend in \drvm\ vs. \logg\ can be best appreciated in the cumulative histograms shown in Figure~\ref{DRVMvloggHist}, which
highlight the tail of the \drvm\ distribution above a few \kms. Here we have divided our main sample of non-RC stars in eight
\logg\ bins, one for subgiant and MS stars (\logg$>3.25$), and seven for RGB stars down to \logg$\sim0$. These bins have roughly
the same number of stars ($\sim$8000), except for the MS/subgiant bin, which has $\sim$19000 (see Table~\ref{tab:NSystems}). The
MS/subgiant bin could in principle be resolved into smaller bins, but given the known limitations of the ASPCAP pipeline for high
\logg\ stars, we felt this was not justified. The histograms in Figure \ref{DRVMvloggHist} show that there is a clear attrition of
high \drvm\ values as stars climb the RGB, and that the distribution of \drvm\ in the RC closely resembles that near the tip of
the RGB.

\subsection{From periods and eccentricities to \drvm}
\label{sec:from-peri-eccentr}

The highest value of \drvm\ in a large sample of stars will correspond to the edge-on ($i=90^{\circ}$) binary systems
with the shortest possible orbital periods. This should be the critical period for Roche Lobe Overflow (RLOF),

\begin{equation}
P_{crit} = 2 \pi \mathcal{R}(q) \sqrt{\frac{R^3}{G M}} \label{eq:1}
\end{equation}

where $\mathcal{R}$(q) is the ratio between the radius of the Roche Lobe and the orbital separation \citep[which can be
approximated by 0.38 for q=1,][]{Eggleton1983}, and $M$ and $R$ are the mass and radius of the photometric primary (i.e., the star
that contributes most of the flux in the APOGEE bands). For any period $P$, the peak-to-peak amplitude of the RV curve of an
edge-on orbit, \drvpp, is twice the semiamplitude $K$,

\begin{equation}
\Delta RV_{pp} = 2 K = \frac{2}{\sqrt{1-e^2}}\left( \frac{\pi GM}{2P} \right) ^{1/3} \label{eq:2}
\end{equation}

where $e$ is the eccentricity. It follows from equations \eqref{eq:1} and \eqref{eq:2} that the value of \drvpp\ at $P_{crit}$ for
a circular orbit with $q=1$ is uniquely determined by the mass and radius of the primary. For stars whose effective gravity $g$ is
measured directly, as our APOGEE targets, these equations can be combined into a simple expression,

\begin{equation}
\Delta RV_{pp,P_{crit}} = 0.87 (GMg)^{1/4}. \label{eq:3}
\end{equation}

\begin{figure}
\centering
\includegraphics[width=0.48\textwidth]{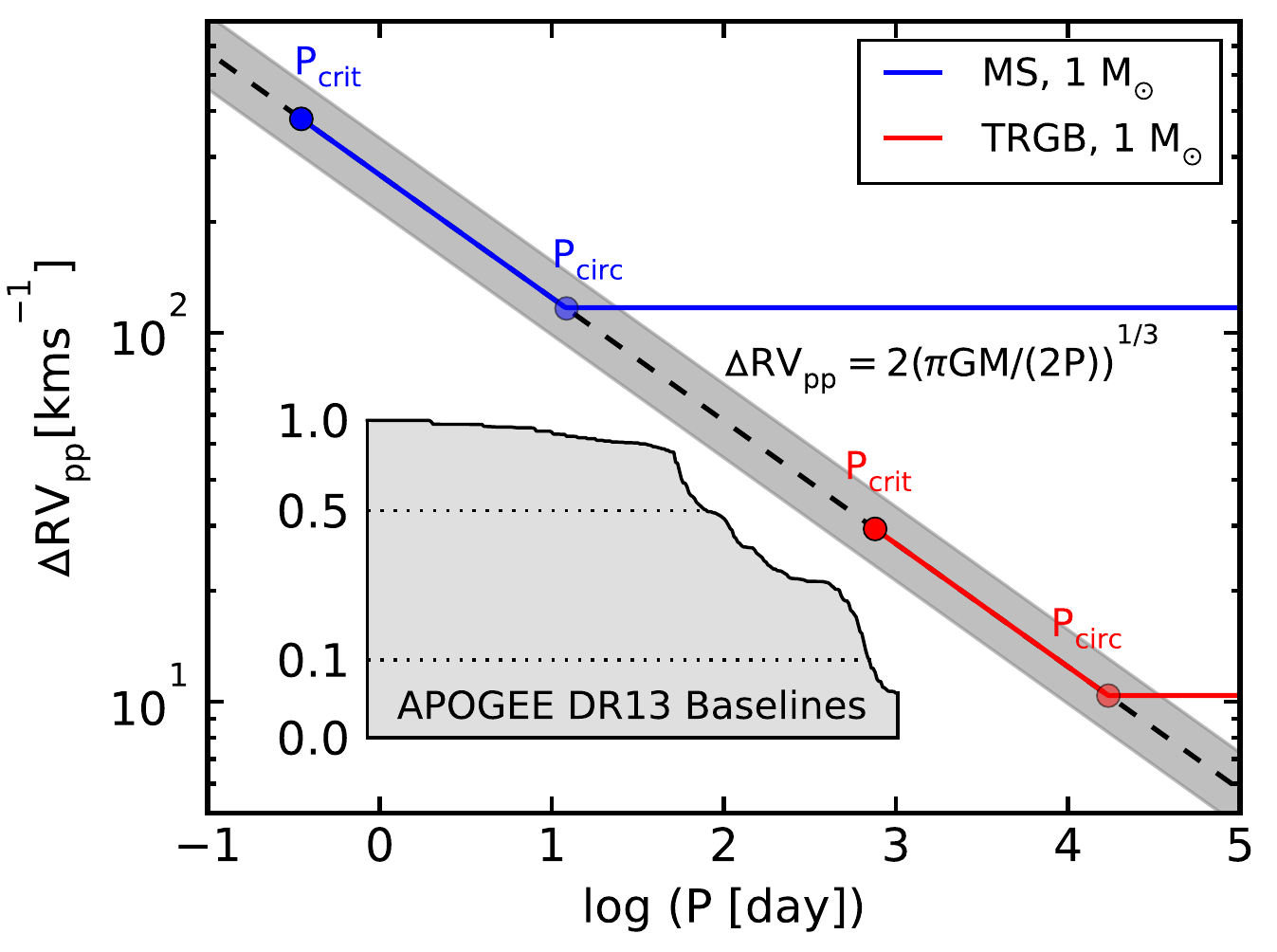}
\caption{Relationship between $P$ and \drvpp\ for edge-on, q=1 systems with circular orbits and 1 \msun\ primaries (dashed
  black line). The range of \drvpp\ for systems with primaries between 0.5 and 2 \msun\ is shown by the gray band. The
  maximum value of \drvpp\ for eccentric systems with 1 \msun\ primaries is shown by the solid colored lines: blue for the MS and
  red for the tip of the RGB (\logg=0), with the values of $P_{crit}$ (from equation~\eqref{eq:1}) and $P_{circ}$ (from
  observations and the theory of \cite{Verbunt1995} applied to the MIST models of \citealt{Choi2016}) marked by symbols. The inset shows
  the fraction of APOGEE targets in our sample that have temporal baselines larger than half a given period, and therefore could
  theoretically probe the full range of the RV curve.}
\label{DRVMPeriod}
\end{figure}

Therefore, a theoretical maximum for \drvpp\ can be estimated for any measured \logg\ just by assuming a mass, as shown by the
solid lines overlaid on Figure~\ref{DRVMvlogg}. For $M$=1 \msun, $P_{crit}$ is 0.35 days in the MS (\logg$\sim$4.5) and 2.1 years
at the tip of the RGB (\logg$\sim$0), and \drvpp\ should vary between 400 and 30 \kms, as shown schematically in
Figure~\ref{DRVMPeriod}. These theoretical maxima compare well with the \drvm\ measurements in Figures~\ref{DRVMvlogg} and
\ref{DRVMvloggHist}, allowing for the fact that the observed systems have a distribution of primary masses, mass ratios, and
inclinations, and their RV curves are sparsely sampled by APOGEE.

\begin{figure}
\centering
\includegraphics[width=0.48\textwidth]{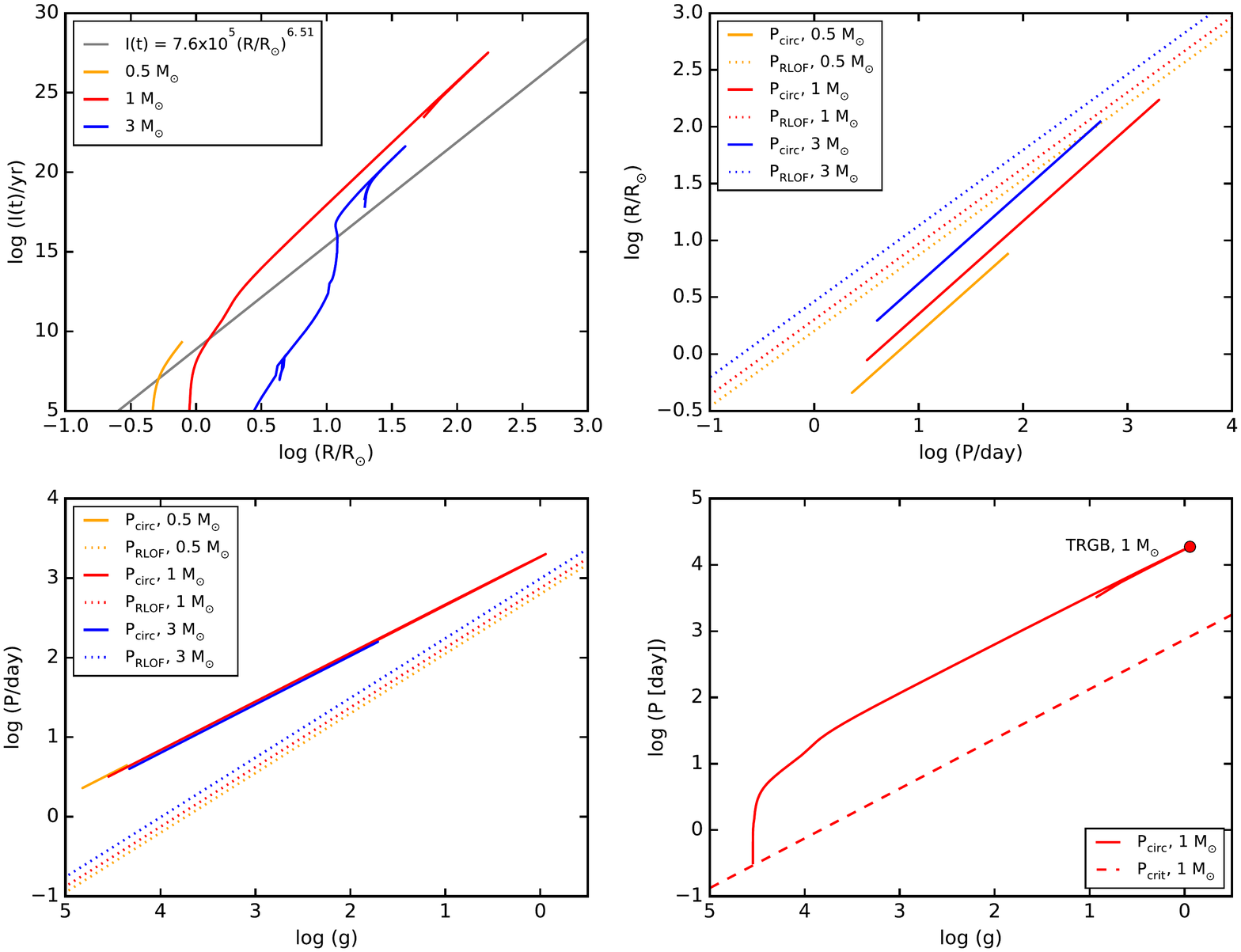}
\caption{Values of $P_{crit}$ and $P_{circ}$ as a function of \logg\ for 1 \msun\ stars. The values for $P_{crit}$ are calculated
  from Equation~\eqref{eq:1}. The values for $P_{circ}$ are calculated using the theory of \cite{Verbunt1995}, with the depth of
  the convective envelope taken from the 1 \msun, [Fe/H]=0 model from \cite{Choi2016}. }
\label{MIST_Pcirc}
\vspace{0.25 cm}
\end{figure}

Orbital eccentricities will also affect \drvm, even though tidal interactions will circularize the orbits with
the shortest periods, i.e., those contributing the largest \drvm\ at any given \logg, on timescales that are comparable to the
evolutionary timescale on the RGB \citep{Meibom2005}. For Sun-like MS stars, \cite{Raghavan2010} measured a circularization period
of $P_{circ}\sim$12 days ($\log{P_{circ}}\sim$1.1). Unfortunately, there is no comprehensive observational study of $P_{circ}$ in
RGB stars with varying \logg. From the theory of tides, \cite{Verbunt1995} derive the value of $P_{circ}$ as a function of the
depth of the convective envelope in the RGB. Taking the depth of the convective envelope from the 1 \msun\ model of solar
metallicity by \cite{Choi2016}, the expressions in \cite{Verbunt1995} predict a steady increase of $P_{circ}$ with decreasing
\logg, with $\log{P_{circ}}\sim$4.2 at the tip of the RGB (Figure~\ref{MIST_Pcirc}). This provides a baseline theoretical scenario
for orbital circularization in the RGB. The maximum eccentricity allowed at any period above $P_{circ}$ is limited by angular
momentum conservation, and can be approximated as

\begin{equation}
e_{max} = \sqrt{1.0-\left( \frac{P_{circ}}{P} \right)^{2/3}} \label{eq:4}
\end{equation}

\citep{Mazeh2008}, which has been shown to be in good agreement with \textit{Kepler} observations of `heartbeat stars'
\citep{Shporer2016}. Note that the period exponents in equations \eqref{eq:2} and \eqref{eq:4} cancel out, and therefore the
maximum \drvpp\ remains constant for periods above $P_{circ}$ (horizontal blue and red lines in Figure~\ref{DRVMPeriod}). In
practice, however, \drvm\ values close to this theoretical upper limit for highly eccentric systems are hard to observe at periods
much longer than $P_{circ}$ because (1) there is a limit to the temporal baseline that can be probed in any RV survey, (2) systems
with $e \sim e_{max}$ are rare, and (3) at high eccentricities it becomes increasingly difficult to capture the full dynamic range
of RV with a sparse sampling, since most of the variation happens in a brief time interval close to periastron. In any case,
orbital eccentricities will break down the simple relationship between $P$ and \drvpp\ shown in Equation~\eqref{eq:2} for periods
longer than $P_{circ}$.

To help visualize the effects of the APOGEE sampling on the distribution of \drvm, we have added an inset to
Figure~\ref{DRVMPeriod} that shows the fraction of stars that could in principle probe the full value of \drvpp, because they have
temporal baselines longer than half a given period. The temporal baselines of individual APOGEE targets range between 0.8 and 1043
days, with a median of 40 days. The fraction of targets that can fully sample the maximum RV range as a function of period falls
rapidly above \logp$\sim$1.7, but it remains above 10\% at \logp$\sim$2.8, which is the value of $P_{crit}$ at the tip of the RGB
for systems with 1 \msun\ primaries. There are no large systematic variations of the distribution of temporal baselines or the
number of visits with either \logg\ or metallicity in our APOGEE targets.

\subsection{Measurement errors and multiple systems}

Even though the APOGEE data reduction pipeline reports RV errors below 0.1 \kms\ \citep{Nidever2015}, \cite{Cottaar2014} found
evidence that these errors might be underestimated by as much as a factor $\sim$3. The \drvm\ distributions measured by APOGEE
also indicate that either the average RV errors are larger than reported by the pipeline or there is an additional source of scatter in
the individual RV measurements.

\begin{figure*}
\centering
\includegraphics[width=0.98\textwidth]{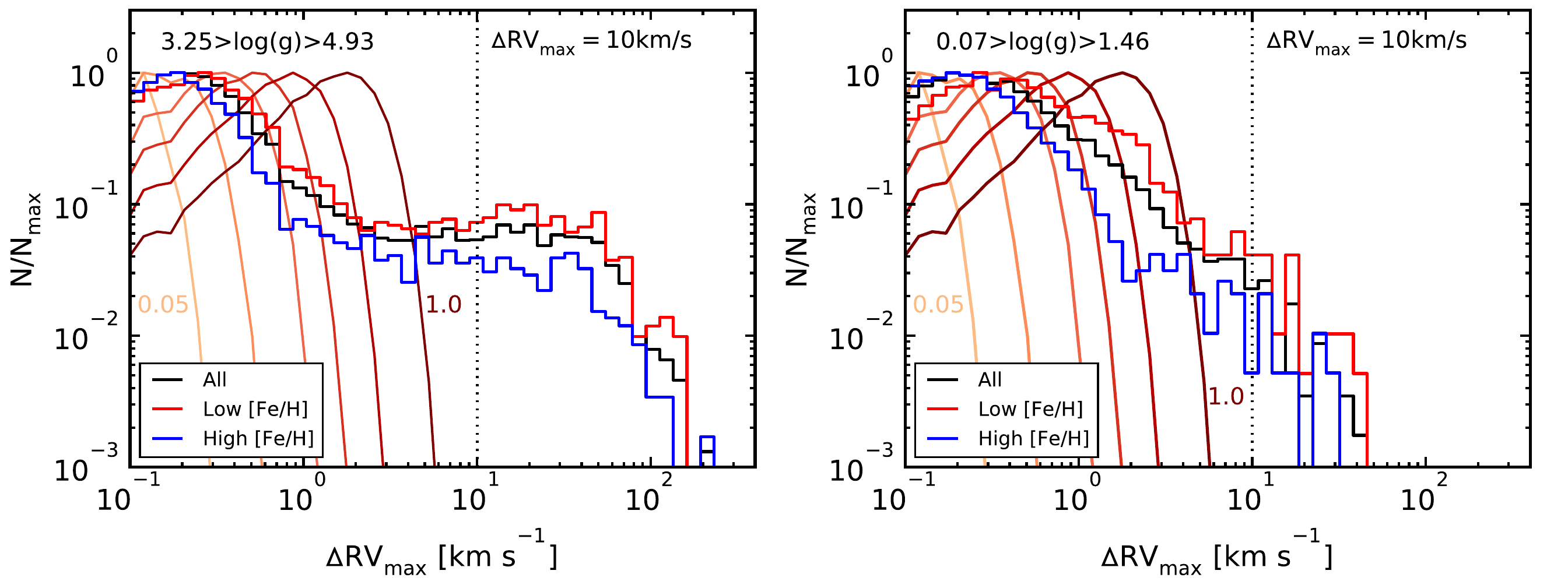}
\caption{Logarithmic \drvm\ distributions for the highest (left) and lowest (right) \logg\ subsamples. For each subsample, we show
  all stars with the black histogram, as well as the top tercile in [Fe/H] (blue histogram) and the bottom tercile in [Fe/H] (red
  histogram). The orange-to-red plots represent simulations of \drvm\ measurements without binaries, using the APOGEE epochs to sample
  constant RV curves with Gaussian RV errors with $\sigma_{RVerr}$ = 0.05 (lightest shade), 0.1, 0.2, 0.3, 0.5, and 1.0 (darkest
  shade) \kms.}
\label{DRVM_Core_tail}
\vspace{0.25 cm}
\end{figure*}

In a survey with a large number of sparsely sampled RV curves, the \drvm\ distribution usually has a core of low values dominated
by measurement errors and a tail of high values from \textit{bona fide} RV variables \citep{Maoz2012}. Even without a detailed
understanding of the RV errors, the transition between tail and core can often be determined empirically, and objects above the
transition can be identified as \textit{bona fide} multiple systems \citep[see][]{Badenes2012,Maoz2016}. In the left panel of
Figure~\ref{DRVM_Core_tail} we show the \drvm\ distribution in our APOGEE MS/Subgiant sample, as well as that of the stars in the
highest and lowest [Fe/H] terciles within this sample. The core/tail transition can be clearly seen at $\sim$0.7 \kms, with the
high and low metallicity terciles having somewhat lower and larger transition values, respectively. Also shown in
Figure~\ref{DRVM_Core_tail} are models where a Gaussian distribution of RV errors is sampled with the APOGEE visits. A comparison
to these models indicates a value of $\sigma_{RVerr}$ close to 0.2 \kms\ for the MS/Subgiant sample. These \drvm\
distributions, and the inferred RV errors, are representative of the majority of our \logg\ samples. The one exception is shown in
the right panel of Figure~\ref{DRVM_Core_tail}. The APOGEE targets with the lowest \logg\ values (0.07$>$\logg$>$1.46) have
noticeably broader \drvm\ cores, with a core/tail transition as high as $\sim$5.0 \kms\ for the stars in the lowest [Fe/H]
tercile. Some of these stars might have RV errors as high as $\sigma_{RVerr} \sim 1$ \kms.

This behavior of the RV errors is qualitatively consistent with the properties of the cross-correlation procedure used by the
APOGEE data reduction pipeline to measure RVs, which is sensitive to parameters that affect the number, depth and broadness of
spectral lines, like \logg\ and [Fe/H]. Quantitatively, the extent of our \drvm\ cores seems to require larger RV errors than
those reported by the pipeline \citep[see Section 8.3 in][]{Nidever2015}. This might be due to a systematic underestimation of the
RV errors by the pipeline or to the presence of a non-Gaussian tail of RV errors, but a more likely explanation is RV jitter,
which can introduce a scatter of up to a few \kms\ in the spectra of RGB stars, especially at low \logg\
\citep{Carney2003,Hekker2008}. To ensure that RV errors, whatever their origin, will not affect our analysis, and that we will be
able to compare stars of different \logg\ and [Fe/H] without biases, we define a conservative threshold of \drvm=10 \kms\ to
identify a system as a \textit{bona fide} multiple. This value is well within the tail of the \drvm\ distribution even in the most
unfavorable cases shown in Figure~\ref{DRVM_Core_tail}, but is low enough to yield a total of 2088 \textit{bona fide} multiple
systems, with enough systems in each \logg\ sample for statistical analysis (see Table~\ref{tab:NSystems}). From
Equation~\eqref{eq:2}, the minimum period required to get \drvm=10 \kms\ is \logp =4.3 ($2 \times 10^4$ days, or 54 yr), which is
close to the expected circularization period for Sun-like stars at the tip of the RGB (Figure~\ref{MIST_Pcirc}). However, the
temporal baselines in APOGEE cannot probe such long periods (see Figure~\ref{DRVMPeriod}). The longest period we are sensitive to
with this \drvm\ threshold is that which can produce a RV shift of at least 10 \kms\ in the longest baselines available in APOGEE
($\sim10^3$ days), which is about \logp$\sim$3.3, or 5.5 yr.

\subsection{Physical Interpretation: Monte Carlo models of \drvm}
\label{sec:phys-interpr}

The relationship between \drvm\ and \logg\ shown in Figures~\ref{DRVMvlogg} and \ref{DRVMvloggHist} can be understood
qualitatively through the equations introduced in Section \ref{sec:from-peri-eccentr} and the interplay between stellar
multiplicity and stellar evolution. After $\sim$1 \msun\ primaries exhaust H in their cores and leave the MS, they climb the RGB
and their \logg\ drops from $\sim$4.5 to $\sim$0 as their radii increase from $\sim$1 to $\sim$170 $R_{\odot}$. For those in
multiple systems, the maximum value of \drvm\ allowed for $P=P_{crit}$ drops from $\sim$400 \kms\ to $\sim$30 \kms\ (Equation
\eqref{eq:3}). Because we cannot find any multiple systems in APOGEE with \drvm\ values above these limits (i.e., to the right of
the solid lines in Figure~\ref{DRVMvlogg}), all systems with $P<P_{crit}$ must have been removed from the sample by some efficient
process. This removal of short $P$ systems also results in a lower number of stars observed at all values of \drvm, as seen in
Figure~\ref{DRVMvloggHist}, due to the projection effect of random orbital inclinations (multiply Equation~\eqref{eq:3} by a
factor $\sin{i}$ that is randomly distributed between 0 and 1). After core He ignition, the stars settle on the RC and their radii
decrease again to $\sim$10 $R_{\odot}$, but their \drvm\ distribution remains similar to that of the larger stars at the tip of
the RGB, because their short period companions have already been removed during shell H burning.

A detailed quantitative evaluation of this scenario, including constraints on multiplicity fractions and period distributions,
would require forward modeling of the APOGEE \drvm\ measurements within a hierarchical Bayesian scheme, taking into account all
the relevant stellar properties and the details of the mass distributions and tidal interactions for the entire sample \citep[e.g,
see][]{Maoz2012,Walker2015c}. We leave that analysis for future work, and here we examine the main physical implications of our
observations using a simpler method.

We generate artificial populations of stars that can be sampled with the APOGEE epochs using a Monte Carlo code. Our code assumes
that all photometric primaries are 1 \msun\ (see Figure~\ref{NessMassHist} and accompanying discussion), that the distribution of
mass ratios is flat \citep[a good first approximation for short period companions to Sun-like stars,][]{Moe2017}, and that orbital
inclinations are random (i.e., the distribution of $\cos{i}$ is uniform).  For each run, we choose a MS multiplicity fraction
$f_{m}$\footnote{Although we only consider binary systems, we call this a multiplicity fraction for consistency. In practice, most
  hierarchical multiple systems contain a tight inner binary that is responsible for most of the RV variation
  \citep{Tokovinin2006,Duchene2013}.}  and an effective gravity \logg. We adopt the period distribution of \citet{Raghavan2010}
(see Figure~\ref{RaghPDist} and accompanying discussion), which we truncate at the value of $P_{crit}$ that corresponds to the
chosen \logg\ (Equation \eqref{eq:1}). We assume that all systems with $P<P_{crit}$ have lost their companions and can be
considered single. We calculate $P_{circ}$ from the theory of \cite{Verbunt1995} and the 1 \msun, [Fe/H]=0 model of
\cite{Choi2016} (Figure~\ref{MIST_Pcirc}), and assume that all orbits with $P_{crit}<P \leq P_{circ}$ are circular. For longer
periods, the eccentricity is drawn from a uniform distribution \citep{Moe2017} between 0 and $e_{max}$ (Equation \eqref{eq:4}). We
generate $N$ systems with these parameters, each of which is sampled with the epochs (number of visits and time lags between
visits) from a random APOGEE target, with the orbital phase of the first visit drawn from a uniform distribution between 0 and
$2\pi$. Thus, the Monte Carlo code captures the main physics affecting the values of \drvm\ with only three free parameters:
$f_m$, \logg, and $N$.

\begin{deluxetable*}{lccccc}
  \tablecaption{Systems with \drvm$>$10 \kms \label{tab:NSystems}}
  \tablehead{
    \colhead{} &
    \colhead{\logg} &
    \colhead{Median} &
    \colhead{} &
    \colhead{} &
    \colhead{} \\
    \colhead{Sample} &
    \colhead{Range} &
    \colhead{\logg} &
    \colhead{$N$} &
    \colhead{$N_{ \Delta RV_{max} > 10 km / s}$} &
    \colhead{$N_{ \Delta RV_{max}  > 10 km / s}/N$}
  }
  \startdata
  MS/Subg. & 3.250 $-$ 4.932 & 4.203 &19045 & 1051 & 0.0552 \\
  RGB 6 & 3.009 $-$ 3.250 & 3.133 & 6211 & 200 & 0.0322 \\
  RGB 5 & 2.791 $-$ 3.009 & 2.876 & 8388 & 183 & 0.0218 \\
  RGB 4 & 2.649 $-$ 2.791 & 2.722 & 8385 & 114 & 0.0136 \\
  RGB 3 & 2.339 $-$ 2.649 & 2.524 & 8412 & 158 & 0.0188 \\
  RGB 2 & 1.931 $-$ 2.339 & 2.141 & 8393 & 135 & 0.0161 \\
  RGB 1 & 1.455 $-$ 1.931 & 1.715 & 8378 & 85 & 0.0102 \\
  RGB 0 & 0.069 $-$ 1.455 & 1.176 & 8366 & 62 & 0.0074 \\
  RC & 2.250 $-$ 3.250 & 2.766 & 15667 & 100 & 0.0064
  \enddata
\end{deluxetable*}

\begin{figure}
\centering
\includegraphics[width=0.48\textwidth]{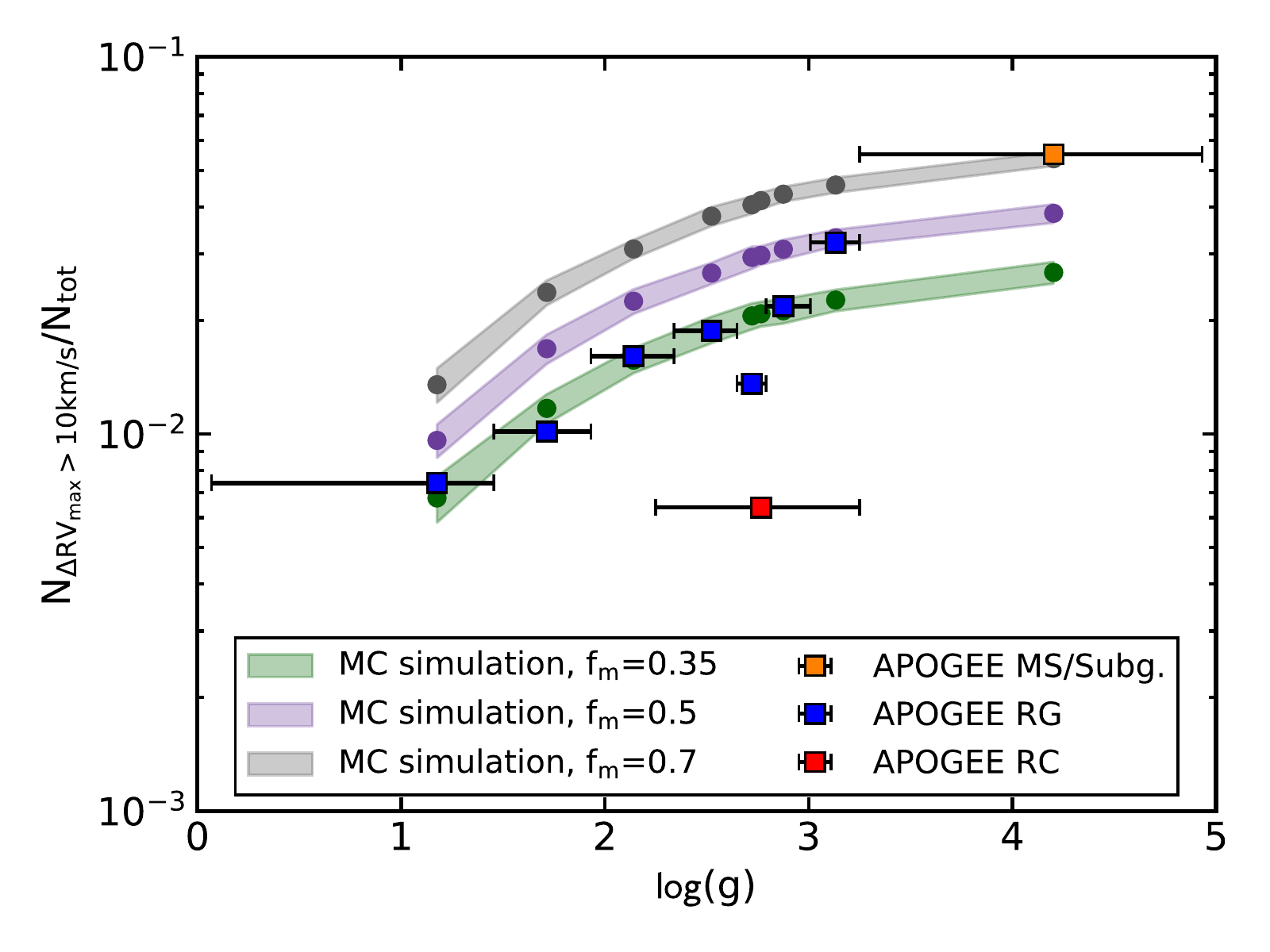}
\caption{Fraction of systems with \drvm$>10$ \kms\ in each \logg\ subsample (orange square for the MS/subgiant sample, and blue
  squares for the RGB samples), as well as the RC catalog (red square). The symbols are placed at the median \logg\ of each
  sample, with the error bars giving the range of \logg\ values. The circles and colored bands represent the results of the
  Monte Carlo simulations described in Section \ref{sec:phys-interpr}, with 1$\sigma$ intervals calculated by running 100
  instances of each simulation with $N$=8300.}
\label{BinFracs_logg}
\vspace{0.25 cm}
\end{figure}

In Figure~\ref{BinFracs_logg} we compare our Monte Carlo simulations with the fraction of targets with \drvm$\geq$10 \kms\
observed by APOGEE in each of the \logg\ samples and in the RC catalog (Table~\ref{tab:NSystems}). The attrition of high \drvm\
(short $P$) systems as stars climb the RGB is clearly seen in the APOGEE data, but the trend seems to break at \logg$\sim$2.7
(sample RGB 4 in Table~\ref{tab:NSystems}). This is the region of the HR diagram where the density peaks of RGB and RC stars overlap
(see Figure~\ref{HR}), which suggests that the lower fraction of high \drvm\ systems in this sample might be due to a
contamination from unidentified RC stars with lower \drvm\ values. This is not surprising, since the RC catalog of \cite{Bovy2014}
is designed to be pure, rather than complete \citep[see][]{Price-Jones2017}.  

Setting aside the issue of the RC contamination, our Monte Carlo simulations reproduce the observed behavior of
$N_{\Delta RV_{max} \geq 10 km s^{-1}}/N$ as a function of \logg\ quite well for RGB stars. The value observed in the RC is close
to that predicted by our simulations at the tip of the RGB, as expected in the framework of our physical model.  In the RGB, we
obtain the best match to the observations for $f_m=0.35$. The RGB sample with the highest \logg\ ($\sim3.1$) and the MS/subgiant
sample seem to require higher values of $f_m$, but these deviations should be interpreted with caution. The APSCAP pipeline was
not designed for stars with high \logg, particularly at low \teff, so the fitted stellar parameters in these samples might be
subject to systematic uncertainties \citep[][note the mismatch between data and models in this region of
Figure~\ref{HR}]{GarciaPerez2016}. Even taking the fitted stellar parameters at face value, it is clear from Figure~\ref{HR} that
these samples must have a primary mass distribution significantly different from the bulk of RGB and RC stars in APOGEE. Our Monte
Carlo code assumes a single value of $f_m$ for all stars, which is not a valid approximation if there is a broad enough range of
primary masses \citep{Lada2006,Moe2017}. Furthermore, the mass ratio distribution, MS period distribution, and tidal
circularization model in our code are only appropriate for 1 \msun\ primaries.  In view of this, the Monte Carlo simulations for
the two highest \logg\ samples might not reflect the underlying properties of the APOGEE targets. We will discuss our measurement
of $f_m$ in more detail in Section \ref{sec:disc}.

\subsection{\drvm\ and metallicity}
\label{sec:effect-metallicity}

The relationship between stellar multiplicity and metallicity is a controversial topic. Some numerical simulations of star
formation show that metallicity should have a strong effect on multiplicity, with lower metallicity clouds showing higher
fragmentation rates and smaller initial separations for binary systems \citep{Machida2008,Machida2009}, but others predict no significant
impact of metallicity for values of [Fe/H] between -1.0 and 0.5 \citep{Bate2014}. Observationally, \cite{Gao2014,Gao2017} found an
inverse correlation between [Fe/H] and multiplicity fraction in field F, G and K-type dwarfs with multiple RVs from the Large Sky
Area Multi-Object Fiber Spectroscopic Telescope \citep[LAMOST,][]{Cui2012}. \cite{Hettinger2015} found the opposing trend in a
smaller sample of F-type dwarfs from SEGUE. However, these studies are based on optical spectra with low resolution (R$\sim$1800),
which makes it hard to confidently measure small RV shifts.

\begin{figure*}
\centering
\includegraphics[width=0.95\textwidth]{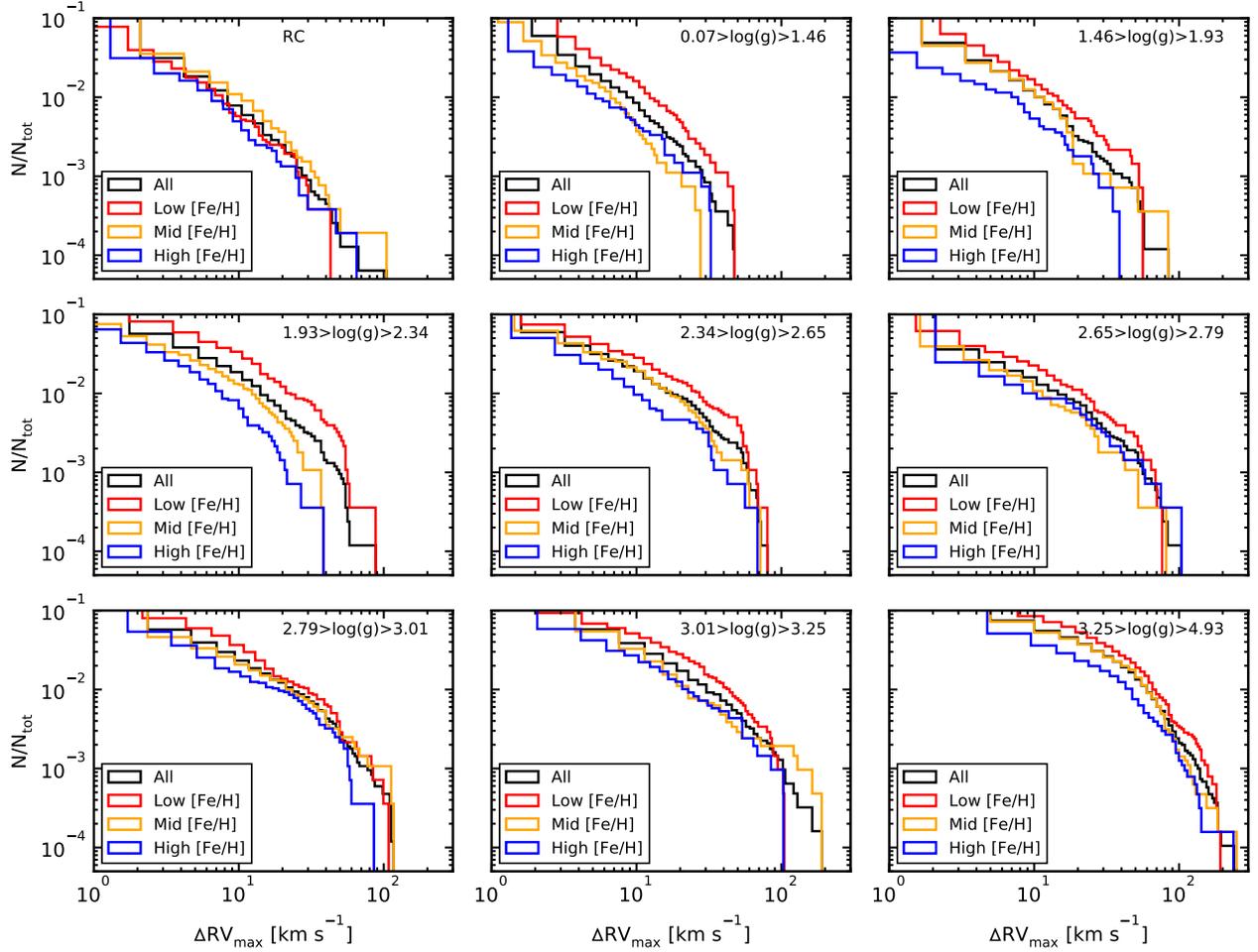}
\caption{Cumulative histograms of \drvm\ for the nine \logg\ subsamples shown in Figure~\ref{DRVMvloggHist}, broken down into
  [Fe/H] terciles.}
\label{DRVMvloggHistFeH}
\end{figure*}

\begin{figure}
\centering
\includegraphics[width=0.48\textwidth]{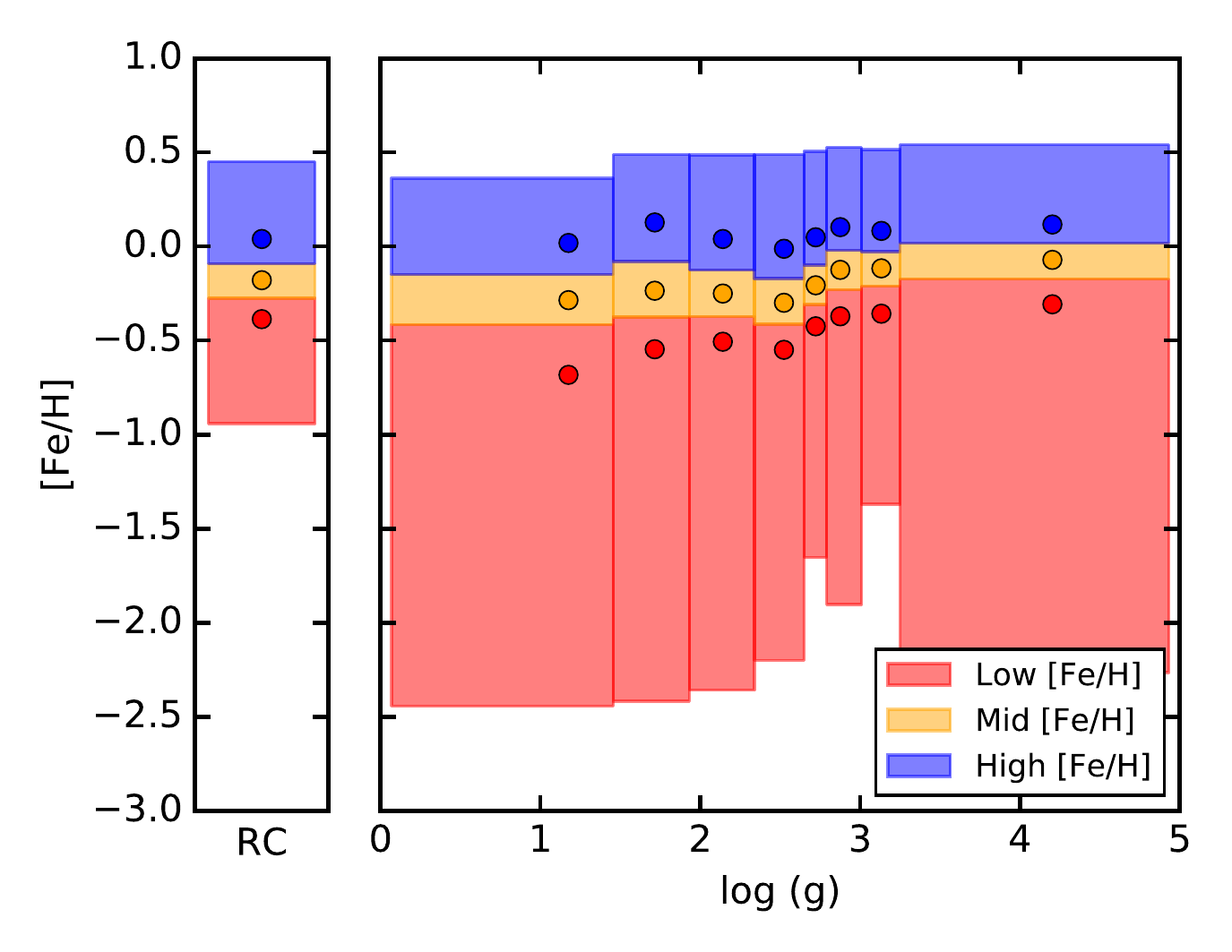}
\caption{Location of the [Fe/H] terciles in each \logg\ subsample, plus the RC. The symbols indicate the median \logg\ of each
  subsample and the median [Fe/H] of each tercile.}
\label{FeH_Terciles}
\vspace{0.25 cm}
\end{figure}

Here we address this topic with the high-resolution IR spectra from APOGEE. In Figure \ref{DRVMvloggHistFeH} we break down the
\drvm\ distribution in each of the samples from Table~\ref{tab:NSystems} into [Fe/H] terciles. The boundaries and median values of
these terciles are shown in Figure~\ref{FeH_Terciles}. There is some variation of the tercile boundaries with \logg, but for the
purpose of this discussion we can consider the low tercile as [Fe/H]$\lesssim-0.5$, and the high tercile as [Fe/H]$\gtrsim0.0$.

\begin{figure}
\centering
\includegraphics[width=0.48\textwidth]{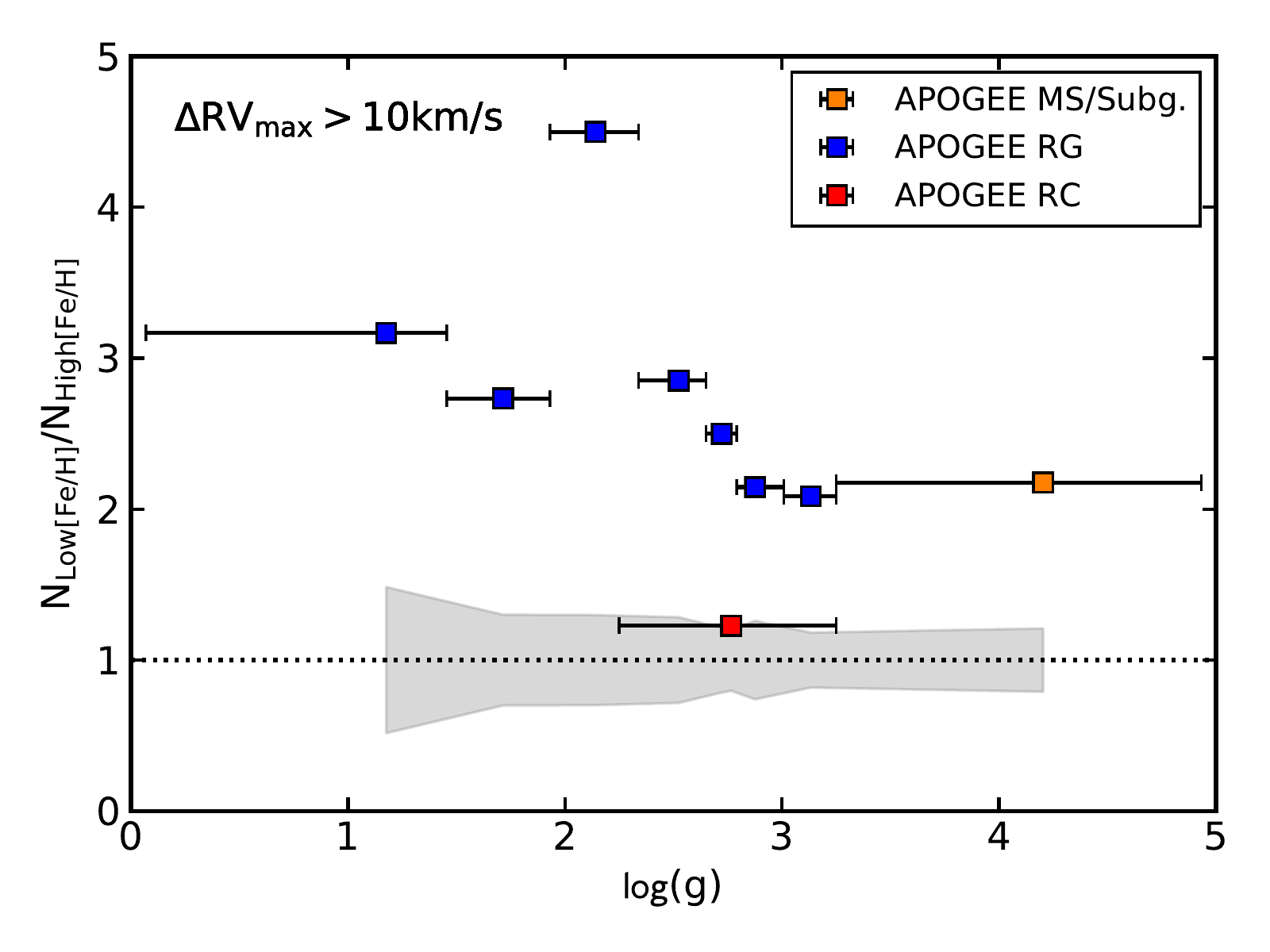}
\caption{Ratio between the fraction of systems with \drvm$>10$ \kms\ in the high and low [Fe/H] terciles of each \logg\ sample, as
  well as the RC catalog.The symbols are placed at the median \logg\ of each sample, with the horizontal error bars giving the
  range of \logg\ values. The shaded grey area around 1 is the 1$\sigma$ probability interval obtained by comparing random
  terciles in our Monte Carlo runs with $N$=8300.}
\label{BinFracs_logg_FeH}
\end{figure}

In all non-RC samples from the MS/subgiants to the tip of the RGB, the \drvm\ distribution in the lowest [Fe/H] tercile is clearly
above that in the highest [Fe/H] tercile. This is true at all values of \drvm\ above our 10 \kms\ threshold, provided there are
enough objects to have small Poisson noise (i.e., $N_{\Delta RV_{max} \geq 10 km s^{-1}}/N $ above a few times $10^{-3}$ for most
of our samples). The ratio between the fraction of systems with \drvm$\geq$10 \kms\ in the high and low [Fe/H] terciles is shown
in Figure \ref{BinFracs_logg_FeH}, together with the 1$\sigma$ probability interval obtained from comparing random terciles in our
Monte Carlo simulations. The ratio is roughly between 2 and 3 for all non-RC samples except the sample at \logg$\sim$2.1, where it
reaches 4.5. The only sample that shows no significant metallicity effect is the RC. This could be related to the fact that the
spread of metallicities in the RC catalog is smaller than in the other APOGEE samples (see Figure~\ref{FeH_Terciles}). Stellar
structure in the RC is itself sensitive to metallicity \citep{Girardi2016}, so very low metallicity He-burning stars
([Fe/H]$\sim$-1) will actually be on the horizontal branch instead of the RC. This introduces a selection effect that leads to a
narrower [Fe/H] distribution, where a multiplicity trend with metallicity can be easily washed out by errors in the APSCAP stellar
parameters.


The most salient features of Figures~\ref{DRVMvloggHistFeH} and~\ref{BinFracs_logg_FeH} can be interpreted in simple terms. It is
clear that low metallicity and high metallicity stars in APOGEE must have different multiplicity characteristics. Either the MS
multiplicity fraction or the MS period distribution (or both), must be metallicity dependent, though the clear segregation at all
values of \drvm\ seen in Figure~\ref{DRVMvloggHistFeH} favors a metallicity dependent multiplicity fraction \citep[see
discussion in Section 3 of][]{Maoz2012}. Provided that the MS period distribution remains roughly lognormal with a sharp
truncation at $P_{crit}$ for RGB stars of all metallicities, the multiplicity fraction of APOGEE targets in the lower metallicity
tercile should be a factor 2-3 higher than in the higher metallicity tercile to reproduce the ratios shown seen in
Figure~\ref{BinFracs_logg_FeH}. If $f_{m}\sim$0.35 for the mean ([Fe/H]$\sim-0.3$), the multiplicity fraction for metal-poor
([Fe/H]$\lesssim-0.5$) stars should be close to 0.46, and that for metal-rich ([Fe/H]$\gtrsim0.0$) stars close to 0.23. The rising
trend in the ratio of high \drvm\ systems for \logg$\lesssim3$ might be related to the shift towards lower metallicity in the
[Fe/H] distributions in that \logg\ range seen in Figure~\ref{FeH_Terciles}.

Although a metallicity dependent multiplicity fraction does provide the simplest explanation for the APOGEE observations, we
emphasize that our simple analysis can only offer a partial glimpse into what is likely a complex situation. For example, it is
well known that \teff, which we have ignored, is strongly correlated with [Fe/H] in the RGB \citep[see Figure 12
in][]{Holtzman2015}, with lower [Fe/H] stars having higher \teff\ at constant \logg. Since [Fe/H] is an intrinsic property of the
star that is set at birth, we have chosen to present our results as a function of [Fe/H], rather than \teff, but see
\cite{Gao2017} for a discussion of the interplay between the two parameters. Another factor that could in principle influence the
\drvm\ distributions is the mass of the primary, but we have found no large systematic differences in the masses determined by
\cite{Ness2016} for the high and low metallicity terciles in our samples, so this effect, if present, must be small.

\section{Discussion}
\label{sec:disc}

\subsection{Multiplicity Fraction}

The best fit value of $f_m\sim0.35$ obtained by comparing our Monte Carlo simulations to the APOGEE observations of RGB stars is
$\sim 60 \%$ lower than the canonical multiplicity fraction for Sun-like MS stars \citep[$0.50 \pm 0.04$ for \logp$<8$, which is
equivalent to 0.56 over the entire period range,][]{Moe2017}. A direct comparison between these two numbers, however, is not
trivial. Our code assumes the \cite{Raghavan2010} period distribution (lognormal, with a peak at \logp$\sim$5.0,
Figure~\ref{RaghPDist}), but the APOGEE baselines are only sensitive to systems in the tail of this distribution, below
\logp$\sim$3.3. Stellar multiplicity surveys usually have a small number of systems with such short periods, so the value of $f_m$
is more uncertain in this period range than the integral over all periods. After correcting the \cite{Raghavan2010} sample for
completeness, \cite{Moe2017} list 32 systems with \logp$<3$, which implies a $\sqrt{N}/N = 18 \%$ uncertainty in the value of
$f_m$ from Poisson statistics alone, much larger than the $8\%$ given for the entire sample. The tidal circularization model in
our Monte Carlo code, which is derived from theory alone, might be inaccurate, and this could impact the number of high \drvm\
systems produced by our simulations. In Monte Carlo runs where the value of $P_{circ}$ shown in Figure~\ref{MIST_Pcirc} is divided
by 2 and 5, we find that the relative increase in $N_{\Delta RV_{max} \geq 10 km s^{-1}}/N$ is $\sim 3\%$ and $\sim 5\%$,
respectively. Finally, \cite{Moe2017} discussed the probability that the photometric primary in a SB1 binary might in fact be the
mass secondary in a post-common envelope system with a white dwarf companion (see their Section 8.3 and Figure 29). The fraction
of short-period (\logp$<3.3$) SB1 systems with white dwarf companions depends on the age of the stellar population, and is
$\sim15\%$ at 1 Gyr and $\sim30\%$ at 10 Gyr. Because the RGB stars in APOGEE have a median age of 4.6 Gyr \citep{Ness2016}, this
fraction should be around $\sim20\%$ for the bulk of our sample, but the impact on high-metallicity and low-metallicity stars will
be different (see below). The combination of the effects we have discussed here ($\sim18\%$ from small sample sizes at short
periods, $\sim5\%$ from uncertainties in the tidal circularization model, and $\sim20\%$ from white dwarf companions) is
comparable to the $\sim 60 \%$ difference between our measurement and that of \cite{Moe2017}. Therefore, we cannot claim that our
measurements of the multiplicity fraction are mutually inconsistent.

\subsection{Metallicity effect}

The metallicity trend shown in Figures~\ref{DRVMvloggHistFeH} and~\ref{BinFracs_logg_FeH} will also be affected by the presence of
white dwarf companions to short-period SB1 binaries. However, the increase of the fraction of white dwarf companions from $15\%$
to $30\%$ between high-metallicity and low-metallicity populations estimated by \cite{Moe2017} cannot explain the factor 2-3
increase in the fraction of high \drvm\ systems seen in APOGEE. In principle, a systematic difference between the masses of stars
in the high and low metallicity terciles of each \logg\ sample could explain our results, but this difference would have to be
very large. If the close (\logp$\lesssim4$) binary fraction scales as $M^{0.7}$ \citep{Moe2017}, the median mass of the stars in
the low metallicity terciles would have to be a factor $\sim$2.6 higher than the median mass of the stars in the high metallicity
terciles to increase the fraction of systems with \drvm$>10$ \kms\ by a factor 2. In our RGB samples, the median masses of stars
in the high and low metallicity terciles measured by \cite{Ness2016} vary by a few tenths of a \msun\ at most.

We conclude that the inverse correlation between metallicity and multiplicity fraction seen in APOGEE must be real. Our result is
in qualitative agreement with the works of \cite{Gao2014} and \cite{Gao2017} using low-resolution spectra of MS stars, and the
findings of \cite{Yuan2015} from photometric studies, but it is in apparent contradiction with the results of \cite{Hettinger2015}
from nearby F-type dwarfs. \cite{Moe2013} determined that the binary fraction at very short periods (\logp$<1.3$) for 10 \msun\
primaries does not vary by more than $\sim30\%$ across the [Fe/H] range between -0.7 and 0.1 by studying early-type eclipsing
binaries in the Milky Way and the Magellanic Clouds. Older studies based on small samples of spectroscopic orbits by
\cite{Latham2002} and \cite{Carney2005} also reported no significant correlation between metallicity and multiplicity. However,
caution must be used again when comparing such different surveys. Because they are not looking at the same stars, and they are not
sensitive to the same periods, it is hard to establish to what degree these observational studies might be mutually inconsistent.
To date, ours is the only study based on high-resolution spectra that span all evolutionary phases between the MS and core He
burning for Sun-like stars in a large fraction of the Milky Way disk, with a sample size orders of magnitude larger than any
previous attempt. The recent re-analysis of the MS spectra in APOGEE by \cite{ElBadry2017} has provided an independent
confirmation of the inverse correlation between metallicity and multiplicity that we report here.

A detailed interpretation of the metallicity effect we have discovered is outside the scope of the present work. Naively, one
might say that our results support the conclusions of \cite{Machida2008} and \cite{Machida2009} regarding the impact of metallicity on star
formation, and that studies where metallicity has little or no effect on stellar multiplicity at birth can be ruled out.
\citep[e.g.][]{Bate2014}. These multiplicity statistics set `at birth' will be subject to dynamical evolution through orbital
capture and disruption, which can be quite different in different environments such as the thin disk, the thick disk, and the
halo. However, numerical simulations imply that the hard binaries we study here (\logp$\leq$3.3) should be relatively unaffected
by this \citep{Kroupa2001,Goodwin2005,Kroupa2011}. More complex processes involving stellar mergers can also affect the
relationship between multiplicity, metallicity, and stellar age \citep[e.g.,][]{Jofre2016,Izzard2018}. 

Regardless of its origin, a metallicity-dependent multiplicity fraction has profound implications for the
rates of interacting binaries like Type Ia supernovae and neutron star mergers, which are observed in host galaxies with a wide
range of redshifts and therefore metallicities \citep{Zahid2013}. Binary population sythesis calculations for these objects often
assume that the multiplicity statistics in the MS (i.e., the initial conditions) are not metallicity dependent
\citep[e.g.][]{Claeys2014,deMink2015}, an assumption that should be revised in light of our results. The dynamical stability of
circumbinary planets will also be affected by a higher fraction of short-period binaries at low metallicities \citep{Jaime2014},
which could be related to the well established paucity of planets around low metallicity stars \citep{Johnson2010}.

\subsubsection{Future Prospects}

For the first time, APOGEE has offered us a panoramic view of the interplay between stellar multiplicity and stellar evolution
across the Milky Way as stars leave the MS, climb the RGB, and settle into the RC after He ignition. This interplay had been
previously constrained by indirect means \citep[e.g., stellar rotation and asteroseismology,][]{Tayar2015}, but our analysis of
the sparsely sampled RV curves from APOGEE allows us to examine it directly and with a large statistical sample. The attrition of
short-period companions as stars undergo RLOF during this process must be rapid and efficient, because the APOGEE \drvm\
measurements are consistent with no systems with $P<P_{crit}$.  Future work on this data set will allow us to quantify this
process and estimate a rate of RLOF events due to unstable mass transfer during H-shell burning in the disk of the Milky Way. The
fate of the systems that are removed from the APOGEE sample is a matter of considerable interest. For mass donors with convective
envelopes, as the APOGEE RGB stars, mass transfer is unstable over a large range in mass ratios \citep{Pavlovskii2015}, and the
outcome should be a common envelope episode in the vast majority of cases. A few of these episodes have now been observed as
optical transients like V838 Mon and V1309 Sco \citep{Ivanova2013}, allowing to constrain the Galactic rate of stellar mergers to
$\sim$0.5 yr$^{-1}$, albeit with large error bars \citep{Kochanek2014}. The few epoch APOGEE spectra can shed light on the fraction of
common envelope episodes that are associated with such transients.

\section{Conclusions}
\label{sec:concl}

We have presented a statistical study of stellar multiplicity that takes advantage of the high resolution and exceptional depth
achieved by the APOGEE IR spectrographs. We defined a sample of $\sim$90,000 stars in the field of the Milky Way with two or
more high-quality RV measurements from APOGEE, spanning the evolutionary stages between the MS and core He burning. We used the
maximum RV shift in these stars, \drvm\ as a figure of merit to study stellar multiplicity, and we established that stars with
\drvm$\geq$ 10 \kms\ are \textit{bona fide} binaries, unaffected by measurement errors. Given the distribution of APOGEE temporal
baselines, we are sensitive to systems with with \logp$\leq$3.3 (5.5 yr). We found a strong correlation between the maximum value
of \drvm\ and \logg, with MS stars having \drvm\ as high as $\sim$300 \kms\ and stars close to the tip of the RGB only going as
high as $\sim$30 \kms. This effect is accompanied by an attrition of \textit{bona fide} binaries at all values of \drvm\ as stars
climb up the RGB. The \drvm\ distribution of core He-burning RC stars is similar to that of stars close to the tip of the
RGB. Finally, we found a strong correlation between metallicity and \drvm, with metal-poor stars in APOGEE ([Fe/H]$\lesssim-0.5$)
being more likely to have any \drvm\ values above our 10 \kms\ threshold than metal-rich stars ([Fe/H]$\gtrsim0.5$).

We used simple Monte Carlo simulations to interpret these observations, and found that the attrition of high \drvm\ systems with
decreasing \logg\ is consistent with the sharp truncation of a lognormal MS period distribution like that measured by
\cite{Raghavan2010} for Sun-like dwarfs in the Solar neighborhood at the critical period for RLOF for each \logg. For the RGB
stars with M$\sim$1 \msun\ that form the bulk of the APOGEE sample and are best characterized by the APSCAP pipeline, we find that
a multiplicity fraction of 0.35 matches the observations quite well. The correlation between \drvm\ and metallicity can be
explained with a metallicity-dependent multiplicity fraction, with metal-poor stars having a factor 2-3 more close companions than
metal-rich stars. This has profound implications for the formation rates of interacting binaries observed by astronomical
transient surveys and gravitational wave detectors, as well as the habitability of circumbinary planets.

\paragraph{Software:} Ipython \citep{Perez2007}, Astropy \citep{Robitaille2013}, Matplotlib \citep{Hunter2007}.

\acknowledgments Jieun Choi kindly made details of her MIST models available to us upon request. Robert Mathieu made useful
suggestions that helped improve this manuscript.  This research was supported in part by Scialog Scholar grant 24215 from the
Research Corporation. C.A.P. is thankful to the Spanish MINECO for funding under grant AYA2014-56359-P. T.C.B. acknowledges
partial support from grant PHY 14-30152: Physics Frontier Center/JINA Center for the Evolution of the Elements (JINA-CEE), awarded
by the US National Science Foundation. This research has made use of NASA's Astrophysics Data System.

Funding for the Sloan Digital Sky Survey IV has been provided by the Alfred P. Sloan Foundation, the U.S. Department of Energy Office of Science, and the Participating Institutions. SDSS-IV acknowledges
support and resources from the Center for High-Performance Computing at
the University of Utah. The SDSS web site is www.sdss.org.

SDSS-IV is managed by the Astrophysical Research Consortium for the 
Participating Institutions of the SDSS Collaboration including the 
Brazilian Participation Group, the Carnegie Institution for Science, 
Carnegie Mellon University, the Chilean Participation Group, the French Participation Group, Harvard-Smithsonian Center for Astrophysics, 
Instituto de Astrof\'isica de Canarias, The Johns Hopkins University, 
Kavli Institute for the Physics and Mathematics of the Universe (IPMU) / 
University of Tokyo, Lawrence Berkeley National Laboratory, 
Leibniz Institut f\"ur Astrophysik Potsdam (AIP),  
Max-Planck-Institut f\"ur Astronomie (MPIA Heidelberg), 
Max-Planck-Institut f\"ur Astrophysik (MPA Garching), 
Max-Planck-Institut f\"ur Extraterrestrische Physik (MPE), 
National Astronomical Observatories of China, New Mexico State University, 
New York University, University of Notre Dame, 
Observat\'ario Nacional / MCTI, The Ohio State University, 
Pennsylvania State University, Shanghai Astronomical Observatory, 
United Kingdom Participation Group,
Universidad Nacional Aut\'onoma de M\'exico, University of Arizona, 
University of Colorado Boulder, University of Oxford, University of Portsmouth, 
University of Utah, University of Virginia, University of Washington, University of Wisconsin, 
Vanderbilt University, and Yale University.

\end{document}